\RequirePackage{fix-cm}
\PassOptionsToPackage{cmex10}{amsmath}
\PassOptionsToPackage{pdftex}{graphicx}
\documentclass[a4paper,british]{article}

\usepackage[utf8]{inputenc}
\usepackage[T1]{fontenc}
\usepackage[hang,flushmargin]{footmisc}
\usepackage{dblfloatfix} 
\usepackage{array}
\usepackage{diagbox}

\usepackage{INTERSPEECH2022}

\usepackage[british]{babel}

\usepackage{multirow}
\usepackage{rotating}

\graphicspath{{./figures/}}
\DeclareGraphicsExtensions{.pdf,.png,.jpeg,.jpg}

\usepackage{amsmath}
\usepackage{amssymb}
\usepackage[unicode=true,pdfusetitle,%
 pdfauthor={Cassia Valentini-Botinhao, Manuel Sam Ribeiro, Oliver Watts, Korin Richmond, Gustav Eje Henter},%
 pdfkeywords={preference prediction, deep learning, speech synthesis},%
 bookmarks=true,bookmarksnumbered=true, pdfborder={0 0 0},colorlinks=false,citecolor=blue]{hyperref}

\title{Predicting pairwise preferences between TTS audio stimuli\\{}using parallel ratings data and anti-symmetric twin neural networks}
\name{Cassia Valentini-Botinhao$^{1,2}$, Manuel Sam Ribeiro$^{1^*}$, Oliver Watts$^{2}$,\\ Korin Richmond$^1$, Gustav Eje Henter$^{3}$}
\address{
  $^1$The Centre for Speech Technology Research, University of Edinburgh, UK \\
  $^2$SpeakUnique Ltd., UK \\
  $^3$Division of Speech, Music and Hearing, KTH Royal Institute of Technology, Sweden
  }
\email{\href{mailto:cvbotinh@inf.ed.ac.uk}{cvbotinh@inf.ed.ac.uk}}

\begin{document}

\maketitle
\thispagestyle{plain}

\begin{abstract} 
Automatically predicting the outcome of subjective listening tests is a challenging task. Ratings may vary from person to person even if preferences are consistent across listeners. While previous work has focused on predicting listeners' ratings (mean opinion scores) of individual stimuli, we focus on the simpler task of predicting subjective preference given two speech stimuli for the same text. We propose a model based on anti-symmetric twin neural networks, trained on pairs of waveforms and their corresponding preference scores.
We explore both attention and recurrent neural nets to account for the fact that stimuli in a pair are not time aligned.
To obtain a large training set we convert listeners' ratings from MUSHRA tests to values that reflect how often one stimulus in the pair was rated higher than the other.
Specifically, we evaluate performance on data obtained from twelve MUSHRA evaluations conducted over five years, containing different TTS systems, built from data of different speakers. Our results compare favourably to a state-of-the-art model trained to predict MOS scores.
\end{abstract}

\section{Introduction}

\let\thefootnote\relax\footnotetext{$^*$The author is now at Amazon. All contributions done while at CSTR.}

Human ratings are the gold standard in evaluating speech generation technologies. Participants are asked to listen to speech stimuli and rate them in isolation or in context. Ratings are subjective, varying from person to person, even more so in cases where the listener is given less context and training. 
The most common type of evaluation is the MOS test \cite{itu1996telephone}. MOS scores are collected from listeners hearing each utterance in isolation, relying on the fact that every person has an internal reference for what is highly natural (score 5) and unnatural (score 1). Even when relative preferences are consistent across listeners, MOS values may vary from person to person. 
Paired comparison tests (so-called `AB tests') exhibit a much lower variance in their responses, which in turn makes statistical significance easier to detect \cite{Yolanda02, kiritchenko17}. Although it is possible to retrieve ordinal rankings from AB tests \cite{Penn18}, MOS (and to some extent MUSHRA tests \cite{Mushra}) are presumably more widely used as their results can directly be used to rank systems in a single scale. In a MUSHRA (MUltiple Stimuli with Hidden Reference and Anchor) test, participants are asked to score systems on a scale from 1--100 by listening to stimuli for the same text side-by-side alongside a high quality reference. Unlike MOS tests, where ratings are given in isolation, MUSHRA test participants are performing a multiple comparison test. This design makes it more sensitive to small differences between stimuli than MOS tests.

Obtaining reliable results from subjective tests requires careful design and a considerable number of listeners \cite{wester:listeners:IS2015}. An easier alternative to human evaluation is promised by so-called objective measures. They can be computed for a given waveform stimulus automatically and are deemed objective as they rely on the same computation no matter who performs it. Although each measure is deterministic, different measures might give drastically different results. The accuracy of a metric (how closely it correlates with human ratings) will most likely depend on the particular characteristics of the stimuli (the speaker, the listener, the linguistic content, the quality of the recording, etc). Designing a measure that works well in every condition is a very challenging task. While measures that are based on auditory models fail to generalise to conditions where the model alone cannot explain human scores, data-driven measures struggle on conditions that were unseen during training. 

Most objective measures were in fact designed to assess natural speech that has been corrupted by a telecommunication channel and/or environmental noise (notably PESQ \cite{Rix01}, POLQA \cite{POLQA} and STOI \cite{Taal10}) rather than artificially generated speech. There has been a growing interest however in metrics for synthesised speech (speech generated by text-to-speech (TTS) systems or voice conversion). Most prominent among these are machine learning models trained on paired audio rating data. These include the models in \cite{Norrenbrock12} which uses the data from Blizzard Challenge 2012 (220 datapoints); the model in \cite{Yoshimura16}, trained on data from 6 years of the Blizzard Challenge \cite{king2014measuring} (3,324 datapoints); AutoMOS, trained on an extensive proprietary dataset of human scores (47,320 datapoints) \cite{Patton16}; MOSNet \cite{Lo2019}, first trained on voice conversion data from the Voice Conversion Challenge (13,580 datapoints) and later on both TTS and voice conversion data from one of the tracks of the ASVSpoof Challenge 2019 (number of datapoints not reported) \cite{Williams2020}; and more recently 
the model proposed in \cite{Mittag_2020} trained on 8 years of Blizzard Challenge data, 2 years of Voice Conversion Challenge plus augmented data generated using POLQA (number of datapoints not reported). All these models are trained to predict MOS ratings for a single waveform. Experience shows that predicting MOS is a very challenging task (esp.\ for unseen speakers, listeners and systems in the out-of-domain setting  \cite{huang2022voicemos}), requiring large amounts of training data.

\begin{figure*}[b!]
    \centering
        \includegraphics[clip, trim=0cm 7.5cm 0cm 10.5cm, width=.85\textwidth]{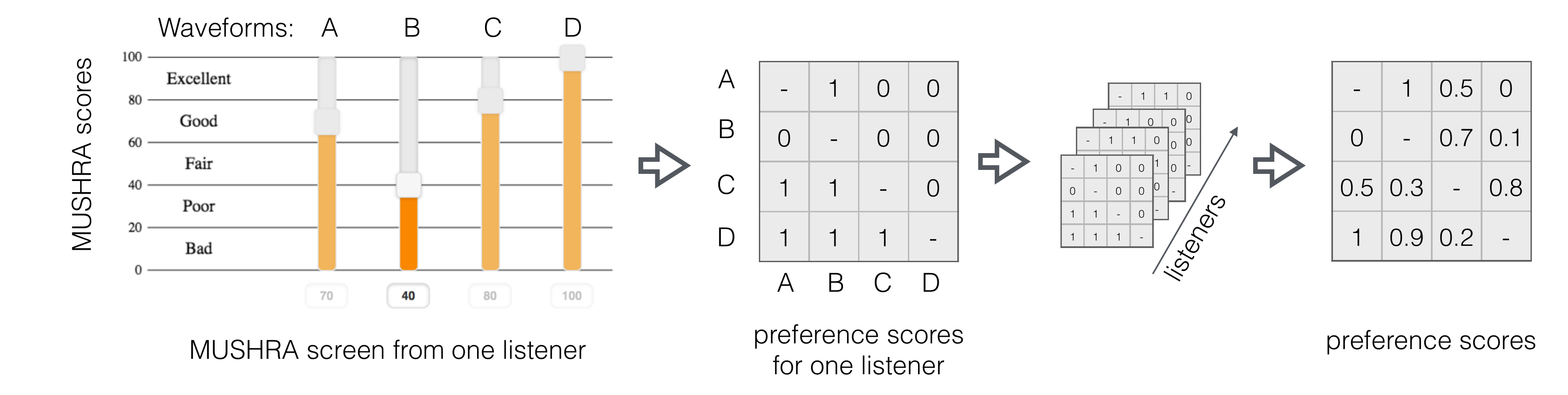}
    \caption{Converting same-screen MUSHRA scores (left) to preference scores (right).}
    \label{fig:fe_mushra_pref}
    \vskip-10pt
\end{figure*}

Although AB scores are more consistent and potentially easier to predict, we know of only one study that has looked into predicting AB scores derived from listening tests \cite{Manocha2020}. 
They proposed a neural network that is trained to predict whether or not two stimuli are perceived as different in listening test responses. 
To obtain a balanced training set, they designed an adaptive listening test that gradually distorts speech using, e.g., additive noise, reverberation and compression. It is not clear how one would gradually distort audio to obtain similar data for text-to-speech. In their more recent work \cite{Manocha2021} the authors incorporate contrastive learning and multi-dimensional learning to encourage the model to learn `content' invariant representations that better generalise to different speakers and sentences.

In this paper we propose \emph{PrefNet}, a neural network trained to predict listener preference given two speech stimuli. In order to create a sizeable set of preference data to train PrefNet we propose converting MUSHRA scores derived from several listening evaluations to pairwise preference scores. We evaluate our system using unseen data (different voices and synthesisers) and compare it to MOSNet \cite{Lo2019}.
Our code, pre-trained models and data are publicly available.$^1$ \footnote{$^1$ \url{ https://github.com/cassiavb/PrefNet}}

\section{Method}
\thispagestyle{empty}

\subsection{Preference scores}

To convert MUSHRA scores into pairwise preference scores we compared the scores of every pair of stimuli belonging to the same MUSHRA screen as shown to the left in Fig.~\ref{fig:fe_mushra_pref}.

This shows which out of two stimuli (different systems speaking the same utterance) a listener preferred. 
The numerical difference between the scores is not taken into account as we assume every participant might use the 0--100 scale differently. The final preference scores were calculated on a per screen basis (i.e., per waveform) as the average of preference scores across participants (right side of Fig.~\ref{fig:fe_mushra_pref}). The strength of the preference is therefore measured in terms of how many participants rated one stimulus above the other, rather than the difference in average score between the two stimuli.

\begin{figure}[t]
    \centering
        \includegraphics[clip, trim=0cm 1.5cm .5cm 0cm, width=.3\textwidth]{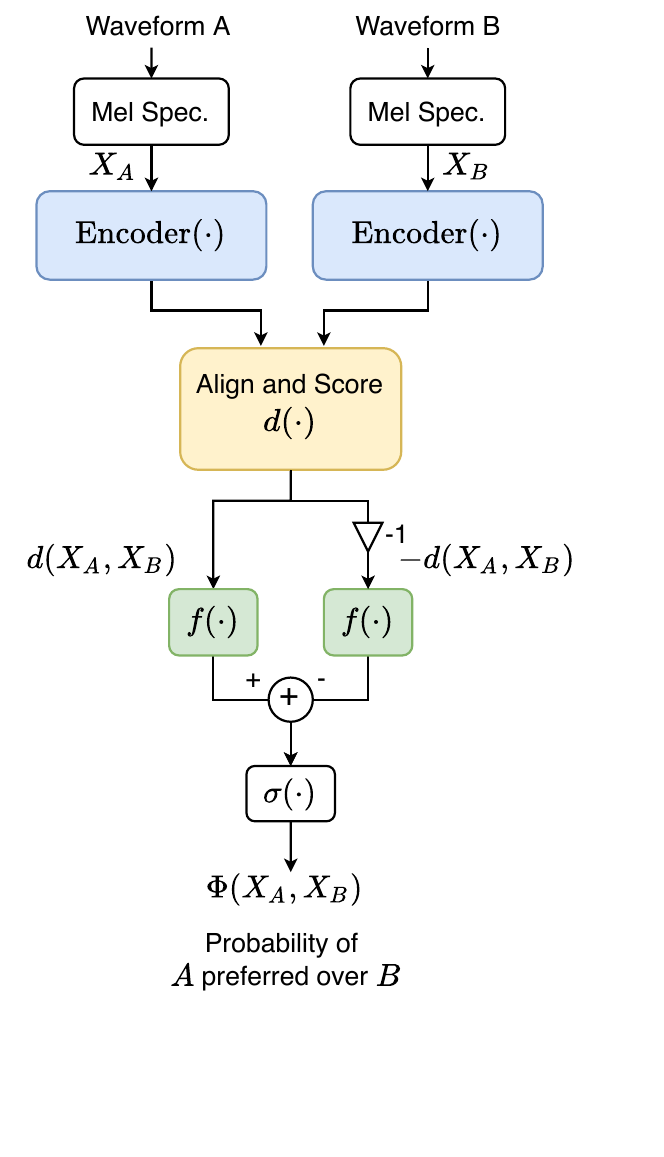}
    \caption{The general architecture of PrefNet. Same-colour boxes denote identical networks with the same weights.}
    \label{fig:fe_models}
    \vskip-10pt
\end{figure}

\subsection{Preference prediction}

Fig.~\ref{fig:fe_models} shows the general architecture of PrefNet. Each input waveform is processed by an encoder network made of a mel spectogram extraction layer (with fixed coefficients) followed by several convolutional layers (which are shared across the different inputs). The variable length encoded representations are then processed by an alignment/scoring mechanism which returns a vector $d(\cdot)$ that essentially subtracts the encoded version of $X_B$ from the encoded version of $X_A$. 
That vector is then processed by a fully connected network $f(\cdot)$ and a sigmoid activation $\sigma(\cdot)$ as follows:
\begin{equation}
    \Phi(X_A,X_B) = \sigma(f(d(X_A,X_B)) - f(-d(X_A,X_B)))
    \label{eq:antisiamese}
\end{equation}
where $X_{A}$ $\in$ $ \mathbb{R} ^{N_A \times 1}$ and $X_{B}$ $\in$ $ \mathbb{R} ^{N_B \times 1}$, $N_A$ and $N_B$ are the number of mel spectogram frames for each input.

PrefNet is trained to predict the (empirical) probability of how often waveform A was preferred over waveform B.
The purpose of Eq.~\eqref{eq:antisiamese} is to ensure that the predicted probabilities are consistent when the input waveforms are swapped: $\Phi(X_B,X_A)=1-\Phi(X_A,X_B)$.
This should lead to more data-efficient modelling, since this property is built in and does not have to be learned.

Architectures where the exact same neural network is applied to multiple inputs are known as \emph{twin neural networks} (or Siamese neural networks). These are often employed in order to learn functions that are symmetric in the input arguments, e.g., similarity scores. The design of PrefNet instead leverages a twin setup to learn a function that is anti-symmetric in the two input stimuli. (Specifically, the output, prior to the final sigmoid, is multiplied by $-$1 if the two inputs change places.) We therefore call our setup an \emph{anti-symmetric twin neural network}.

We propose two versions of PrefNet that vary in terms of how alignment and scoring is performed. 

\begin{table*}[t]
\centering
\setlength{\tabcolsep}{2pt} 
\begin{tabular}{|c|>{\centering\arraybackslash}p{2cm}|>{\centering\arraybackslash}p{1.7cm}|>{\centering\arraybackslash}p{1.7cm}|>{\centering\arraybackslash}p{1.7cm}|>{\centering\arraybackslash}p{1.7cm}|>{\centering\arraybackslash}p{1.7cm}|>{\centering\arraybackslash}p{1.6cm}|>{\centering\arraybackslash}p{1.6cm}|}
\hline
& Large set & Eval$_1$ & Eval$_2$ & Eval$_3$ & Eval$_4$ & Eval$_5$ & Eval$_6$ & Eval$_7$ \\ \hline
Stimuli pairs & 17,047 & 2,160 & 1,440 & 800 & 800 & 800 & 3,300 & 720 \\ \hline
System pairs & 282 & 36 & 36 & 10 & 10 & 10 & 55 & 10 \\ \hline
Systems & 76 & 9 & 9 & 5 & 5 & 5 & 11 & 5 \\ \hline
Speech corpus & Hurricane \cite{Hurricane13} & Blizzard 2013 \cite{blizzard13} & Blizzard 2013 \cite{blizzard13} & Blizzard 2013 \cite{blizzard13} & Blizzard 2013 \cite{blizzard13} & Blizzard 2013 \cite{blizzard13} & LJ Speech \cite{ito2017lj} & Blizzard 2013 \cite{blizzard13} \\ \hline
TTS systems & HTS \cite{hts}, Multisyn \cite{Clark07}, vocoded \cite{Espic17}, Merlin \cite{Merlin16} & DCTTS \cite{Tachibana18} & DCTTS \cite{Tachibana18}& DCTTS \cite{Tachibana18} & DCTTS \cite{Tachibana18} & DCTTS \cite{Tachibana18} & Merlin \cite{Merlin16}, DCTTS \cite{Tachibana18} & DCTTS \cite{Tachibana18}, Tacotron \cite{skerry2018towards} \\
\hline
\end{tabular}
\vskip2pt
\caption{Overview of the data used in the evaluation. The `Large set' contains data from 12 different evaluations.} \label{tab:fe_data}
\vskip-10pt
\end{table*}

\subsubsection{Attention-based scoring}

We investigate an attention-like mechanism \cite{Vaswani17} to align and score the distance between input stimuli. The encoder network of the attention-based model predicts two sets of representations $K$ and $V$ for each input:
\begin{align}
K_{A}, V_{A} = H(\mathrm{Encoder}(X_A))\\
K_{B}, V_{B} = H(\mathrm{Encoder}(X_B))
\end{align}
\noindent where $H(\cdot)$ is an affine layer, $K_{A}$ $\in$ $ \mathbb{R} ^{N_A \times D}$, $K_{B}$ $\in$ $ \mathbb{R} ^{N_B \times D}$, $V_{A}$ $\in$ $ \mathbb{R} ^{N_A \times E}$ and $V_{B}$ $\in$ $ \mathbb{R} ^{N_B \times E}$. 
The attention matrix is then computed as follows:
\begin{equation}
W = \underset{N_A,N_B}{\mathrm{softmax}} ( K_{A} K_{B}^{T} )
\end{equation}
\noindent where $W$ $\in$ $ \mathbb{R} ^{N_A \times N_B}$. Note that the softmax operation is performed across both rows and columns, i.e., all elements of $W$ collectively sum to one.
The distance $d(\cdot) \in \mathbb{R} ^{E \times 1}$ between the two inputs is derived as:
\begin{equation}
    d(X_A,X_B) = \sum_{t_A,t_B=1}^{t_A=N_A,t_B=N_B} ( V_A[t_A,:] - V_B[t_B,:] ) W[t_A,t_B] 
    \text{.}
\end{equation}

\subsubsection{GRU-based scoring}

Rather than comparing the input signals frame by frame, as the attention-based system does, our other system variants process each variable length representation with a shared GRU layer and convert the processed sequence into a fixed length representation. The scoring vector $d(\cdot)$ is then computed as follows:
\begin{equation}
\begin{aligned}
    d(X_A,X_B) = & g ( \mathrm{GRU}(\mathrm{Encoder}(X_A)) ) \\ 
    & - g ( \mathrm{GRU}( \mathrm{Encoder}(X_B)) )
    \label{eq:gru_antisiamese}
\end{aligned}
\end{equation}
\noindent where $d$ $\in$ $ \mathbb{R} ^{E \times 1}$ and $g(\cdot)$ represents the operation that extracts a fixed length representation from the GRU sequence, like from instance its last output.

\section{Evaluation}

\subsection{Data}

Table \ref{tab:fe_data} presents details on the data we used for training and testing. The first column (`Large set') describes data collected from 12 different lab-based MUSHRA tests over a period of 5 years (2014--2019). This includes the evaluations described in \cite{Zhizheng15, Merritt2015, Merritt2016, Watts16, Espic17, Watts18, Zofia19}, among others. The tests involve stimuli of natural and vocoded speech, but with a focus on a variety of TTS systems, including HMM-based systems build with HTS \cite{hts}, unit selection built with Multisyn \cite{Clark07} and early DNN systems built with Merlin \cite{Merlin16}. All systems in these evaluations were created using the same speech dataset, the Hurricane Challenge corpus (male British speaker) \cite{Hurricane13}.

The other columns in Table \ref{tab:fe_data} describe data from more recent lab-based MUSHRA evaluations, containing sequence-to-sequence TTS models like DCTTS \cite{Tachibana18} and Tacotron \cite{skerry2018towards}. Eval$_6$ described in \cite{Watts19} involved voices built with Merlin and DCTTS using the LJ Speech dataset (US female speaker) \cite{ito2017lj}. All other evaluations contained voices based on the Blizzard Challenge 2013 corpus (US female speaker) \cite{blizzard13}.

All audio stimuli were resampled to 16 kHz. We extracted a 64-dimensional log-magnitude mel spectrogram representation using a window of width 512 and 12.5 ms frame hop.

\begin{table}[t]
\setlength{\tabcolsep}{3.5pt} 
\begin{center}
\begin{tabular}{| c | c | c c c c | c | }
\cline{2-7}
\multicolumn{1}{c|}{} & \multicolumn{1}{c}{Test fold:} & Fold 1 & Fold 2 & Fold 3 & Fold 4 & Average\\
 \cline{2-7} \hline
\multirow{6}{*}{\begin{turn}{90}Training folds\end{turn}} 
& \# pairs & 12,281 & 10,237 & 15,647 & 12,976 & - \\ \cline{2-7}
& Attention & 70.0 & 81.3 & 70.8 & 67.5 & 72.4 \\ \cline{2-7}
& GRU$_1$ & 91.2 & 90.4 & 90.6 & 90.0 & 90.5 \\
& GRU$_2$ & 91.9 & 92.3 & 91.8 & 90.4 & 91.6 \\
& GRU$_3$ & {\bf 92.7} & {\bf 92.4} & 92.2 & 90.5 & {\bf 91.9} \\
& GRU$_4$ & 92.5 & 91.8 & {\bf 92.6} & {\bf 90.7} & {\bf 91.9} \\
\hline
\hline
\multirow{6}{*}{\begin{turn}{90}Test fold\end{turn}} 
& \# pairs & 4,766 & 6,810 & 1,400 & 4,071 & - \\ \cline{2-7}
& Attention & 60.0 & 66.6 & 64.8 & 63.6 & 63.8 \\ \cline{2-7}
& GRU$_1$ & 68.7 & 70.8 & 73.6 & 62.2 & 68.8 \\
& GRU$_2$ & 72.9 & 73.4 & 72.4 & 65.2 & 71.0 \\
& GRU$_3$ & 75.3 & {\bf 75.2} & 75.1 & 70.1 & 73.9 \\
& GRU$_4$ & {\bf 76.9} & 74.2 & {\bf 77.1} & {\bf 71.5} & {\bf 74.9} \\
\hline
\end{tabular}
\end{center}
\vskip-5pt
\caption{Prediction accuracy (\%) of different folds (columns) on the training data (top rows) and on the test data (bottom rows).}
\vskip-20pt
\label{tab:fe_results1}
\end{table}

\begin{table*}[t]
\begin{center}
\setlength{\tabcolsep}{4.5pt} 
\begin{tabular}{| c | >{\centering\arraybackslash}p{3.5cm} | c | c | c | c | c | c | c | }
\hline
& \backslashbox[3.5cm]{Train}{Test} & Eval$_1$ & Eval$_2$ & Eval$_3$ & Eval$_4$ & Eval$_5$ & Eval$_6$ & Eval$_7$\\  \cline{1-9}
\multirow{5}{*}{\begin{turn}{90} PrefNet  \end{turn}} 
& Large set & 48.8 / 55.6 & 46.6 / 44.4 & 55.4 / 60.0 & 52.8 / 60.0 & 57.5 / 80.0 & 56.8 / 61.8 & {\bf 68.2} / {\bf 100} \\ \cline{2-9}
& Eval$_1$  & - & {\bf 72.9} / 83.3 & {\bf 75.5} / {\bf 80.0} & {\bf 67.8} / 70.0 & {\bf 84.8} / 90.0 & {\bf 66.3} / 65.5 & 61.4 / 60.0 \\ \cline{2-9}
& Large set and Eval$_1$ (FT) & - & 68.8/ {\bf 88.9} & 69.9 / {\bf 80.0} & 63.2 / {\bf 80.0} & 83.6 / {\bf 100} & 65.5 / 70.9 & 63.5 / 70.0 \\ \cline{2-9}
& Eval$_{1-5}$ & - & - & - & - & - & 62.4 / 61.8 & 65.6 / 50.0 \\ \cline{2-9}
& Large set and Eval$_{1-5}$ (FT) & - & - & - & - & - & 66.1 / {\bf 72.7} & 63.2 / 50.0 \\
\hline
\hline
\multirow{4}{*}{\begin{turn}{90} MOSNet \end{turn}} 
& Voice conversion data (VC) \cite{Lo2019} & 51.2 / 50.0 & 49.2 / 33.3 & 68.6 / {\bf 80.0} & 45.9 / 60.0 & 51.0 / 70.0 & 65.1 / 67.3 & 51.5 / 30.0 \\ \cline{2-9}
& Anti-spoofing data \cite{Williams2020} (TTS, TTS+VC, VC) & {\bf 68.8} / {\bf 86.1} & 61.5 / 69.4 & 54.1 / 60.0 & 44.9 / 40.0 & 65.8 / 80.0 & 43.4 / 40.0 & 31.4 / 20.0 \\ \hline
\end{tabular}
\end{center}
\vskip-7pt
\caption{Prediction accuracy (\%) at stimuli / system level. FT stands for fine-tuned.} \label{tab:fe_results2}
\vskip-20pt
\end{table*}

\subsection{Models}

We performed a hyperparameter sweep on the `Large set' in Table \ref{tab:fe_data} using multifold cross-validation. The best attention-based model we found was made of four convolutional layers of width three (the first layer of dilation three, the others one) over 64 channels ($D$=32 and $E$=32). The encoder of the best GRU-based model was made of two convolutional layers of kernel width 9 over 64 channels. The GRU contained 64 units ($E$=64). For all models the fully connected network $f(\cdot)$ had only one layer.
All models were trained to predict relative preferences (probabilities) using the mean squared error loss (also known as the Brier score). Training ran for 50 epochs with early stopping based on a random selection of 10\% of the training data for validation, using the Adam optimiser with a learning rate of 0.001. The pairs of input sequences were batched according to the size of the longer waveform in each pair. 

We present results of four GRU-based models. 
GRU$_1$ is a variant without the anti-symmetric design whose output is instead given by $\sigma(\, f(g(\mathrm{Encoder}(X_A)) - g(\mathrm{Encoder}(X_B)))\, )$. 
GRU$_2$ is the anti-symmetric version proposed in Eqs.~\eqref{eq:antisiamese} and \eqref{eq:gru_antisiamese} where $g(\cdot)$ extracts the final output of the GRU. 
GRU$_3$ obeys the same anti-symmetric architecture but rather than retrieving the final output, the average representation is obtained.
GRU$_4$ is a bi-directional version of GRU$_3$.

\subsection{Similar conditions: same speaker}

To compare variants of the PrefNet architecture, we performed a multifold cross-validation evaluation using the `Large set' data. We divided the 12 evaluations into four folds ordering the evaluations chronologically, so that different folds are less likely to contain the same systems. (For instance, the first and second fold contained only HMM-based TTS, while only the third and fourth folds had DNN-based voices.) Each architecture was trained four times, varying which fold was held out from training. The prediction accuracy (calculated at a sentence level across all sentences) is reported in Table \ref{tab:fe_results1}. 

We see that accuracy on the training data (upper rows) is relatively high, particularly for the GRU-based models. Performance drops considerably when evaluating on unseen material, i.e., the fold that has been held out (lower rows). 
Overall we can see that the GRU-based models performs better than the attention-based one. GRU$_2$ scores higher than GRU$_1$ indicating that the anti-symmetric architecture is helpful. We observed that introducing it improved not only accuracy, but also model convergence. 
Taking the average over the GRU's output sequence (rather than using its last element) seems to further improve results, as well as using bidirectional GRUs.
The remaining experiments focus on the GRU$_4$ variant of PrefNet.

\subsection{Unseen conditions: different speakers and systems}

To evaluate how well PrefNet predicts the preference between more recent state-of-the-art TTS models we extended the evaluation to the data from evaluations Eval$_{1}$ to Eval$_{7}$, with several different training and test data configurations. The results are presented in Table~\ref{tab:fe_results2}. Rows refer to training material and columns to the test set. We present scores at both stimulus level and system level. For system-level results we calculated the accuracy across all sentences of each system pair.

As a baseline, we compare the various PrefNet results to results obtained from pre-trained MOSNet models, one trained on voice conversion (VC) data \cite{Lo2019} and one trained on both VC and TTS data \cite{Williams2020} including HMM-based systems, Merlin and Tacotron trained on the VCTK corpus (multi-speaker English language dataset) \cite{VCTK}. These represent the state of the art in publicly available models for MOS score prediction on synthetic speech. The MOS scores predicted by MOSNet were converted to pairwise preferences by checking which of two paired stimuli had the higher predicted MOS score.

The results in Table~\ref{tab:fe_results2} show that the model trained on the `Large set' (top row) performs relatively well on Eval$_7$ but not on the other evaluations where the stimulus-level performance is close to chance rate. 
The model trained on Eval$_1$ alone (second row) highlights the importance of representative training data, since it performs well on similar evaluations (i.e., Eval$_{2-5}$) but less so on evaluations with different speakers and systems (Eval$_6$ and Eval$_7$). 
Fine tuning the model trained on the `Large set' with data from Eval$_1$ (third row) seems to improve results on Eval$_{2-5}$ but to the detriment of performance on Eval$_7$.
Training with more quantities of similar data (fourth and fifth row)
does not seem to improve performance on Eval$_6$.
System-level performance is generally better than stimulus-level performance for all systems. System-level accuracy seems to be higher for the fine-tuned models.

The MOSNet models generally perform worse than most PrefNet models, with performance varying considerably depending on the specific test set. Poor performance could to some extent be explained by the fact that these models were trained on data that is potentially more dissimilar to the test set than PrefNet training data.
In several cases, MOSNet performance is below chance rate, with Eval$_7$ being the most challenging to predict. Interestingly, the MOSNet model trained on TTS data was not always the best. A possible explanation is that it used less training data than the VC-only model.

\section{Conclusions}
This paper has introduced PrefNet, an approach to predicting pairwise preferences between synthetic speech stimuli. We demonstrated how data from side-by-side evaluations using numerical ratings can be leveraged to create training data for pairwise preference prediction. We empirically investigated  several architectures and described a design rooted in twin neural nets that ensures consistent pairwise preferences if the order of the inputs is reversed.
Results showed that GRU-based architectures outperformed those using attention to align and score stimulus pairs, and that our anti-symmetric network design also improved accuracy.
PrefNet outperforms MOSNet (a state-of-the-art model trained to predict MOS scores, from which implicit pairwise preferences can be derived) in most conditions but the performance of both models varies considerably depending on training data and test data. The performance of the model trained on the larger set of listening tests (including HMM and DNN systems) improves when data from more recent (and thus more similar) systems are included in training. Future work includes training with both stimulus-level and system-level preferences and increasing the amount of training data by converting MOS (as well as MUSHRA) scores to preferences. 

\vskip5pt

\noindent\textbf{Acknowledgements:}
The work presented in this paper was partially supported by Samsung Electronics Co., Ltd.\ and by the Wallenberg AI, Autonomous Systems and Software Program (WASP) funded by the Knut and Alice Wallenberg Foundation.

\newpage
\bibliographystyle{IEEEtran}

\bibliography{mybib}

\end{document}